\documentclass{iaus}
\usepackage{graphicx}

\title[Spectroscopy of Extragalactic PNe] %% give here short title %%
{The Spectroscopic Properties of Bright Extragalactic Planetary Nebulae}

\author[Richer]   %% give here short author list %%
{Michael G. Richer$^1$}
\affiliation{$^1$Instituto de Astronom\'ia, Universidad Nacional Aut\'onoma de M\'exico\\ Apto. Postal 877, 22800 Ensenada, Baja California, M\'exico \break email: richer@astrosen.unam.mx
}

\pubyear{2006}
\volume{234}  %% insert here IAU Symposium No.
\pagerange{119--126}
\date{?? and in revised form ??}
\jname{Planetary Nebulae in our Galaxy and Beyond}
\editors{M. J. Barlow \& R. H. M\'endez, eds.}
\begin{document}

\maketitle

\begin{abstract}

The properties of bright extragalactic planetary nebulae are reviewed based upon the results of low and high resolution spectroscopy.  It is argued that bright extragalactic planetary nebulae from galaxies (or subsystems) with and without star formation have different distributions of central star temperature and ionization structure.  As regards the chemical compositions, oxygen and neon are generally found to be unchanged as a result of the evolution of the stellar progenitors.  Nitrogen enrichment may occur as a result of the evolution of the progenitors of bright planetary nebulae in all stellar populations, though this enrichment may be (more) random in old stellar populations.  Helium abundances appear to be influenced by the chemical evolution of the host galaxy, with planetary nebulae in dwarf spheroidals having systematically elevated abundances.  Neither the age nor the metallicity of the progenitor stellar population has a strong effect upon the kinematics observed for nebular shells.  Both the range of expansion velocites, 8-28\,km\,s$^{-1}$, and the typical expansion velocity, $\sim 18\,\mathrm{km\,s}^{-1}$, are found to be relatively constant in all galaxies.  On the other hand, bright planetary nebulae in the bulge of M31 have systematically higher expansion velocities than their counterparts in M31's disk.  The expansion velocities show no trend with nebular H$\beta$ luminosity, apart from a lack of large expansion velocities at the highest luminosities (the youngest objects), but appear to correlate with the $5007/\mathrm H\beta$ ratio, at least until this ratio saturates. These results suggest a link between the evolution of the nebular shells and central stars of bright extragalactic planetary nebulae.

\end{abstract}

\firstsection % if your document starts with a section,
\section{Introduction}

Since the last IAU Symposium devoted to planetary nebulae, a considerable body of spectroscopic data for bright extragalactic planetary nebulae has become available (Table \ref{table_datasrc}).  Data not included here, but presented at this conference as well as the recent ones in Garching (in 2004) and Gdansk (in 2005), will substantially increase the data base and allow for the first comparisons of independent data sets, providing a better understanding of the systematic uncertainties in the spectroscopy of these faint objects.  Spectroscopy is typically available at resolutions below $\sim 5\times 10^3$ for the purposes of studying chemical abundances (low resolution) and at resolutions $> 25\times 10^3$ for the purposes of studying the nebular kinematics (high resolution).  Only bright extragalactic planetary nebulae are considered here, namely, those  within $\sim 2.5$\,mag. of the peak of the luminosity function in [O~III]$\lambda$5007.  While all of the selection effects inherent in this selection are not clear, it will favour young planetary nebulae and, at low metallicities, the most oxygen-rich objects in each galaxy (\cite[Dopita \etal\ 1992]{dopitaetal1992}; \cite[Richer \& McCall 1995]{richermccall1995}).  

All of the low resolution spectroscopy available (Table \ref{table_datasrc}) was compiled and analyzed in a uniform fashion.  The data were restricted to techniques permitting background subtraction, since removing the local background is crucial, especially in galaxies with ongoing star formation.  When possible, raw line intensities, uncorrected for reddening, were the starting point, from which reddening-corrected line ratios and ionic abundances were computed.  The ionic abundances are all based upon measured electron temperatures or limits, from the [O~III]$\lambda\lambda4363/5007$ intensity ratio.  When available, the electron density based upon the [S~II]$\lambda\lambda$6716,6731 lines was used.  Otherwise, an electron density of 2000\,cm$^{-3}$ was adopted.  Elemental abundances were derived using the ionization correction factors from \cite{kingsburghbarlow1994}.  The atomic data used is given in \cite{richermccall2006a}.  

In what follows, the stellar populations from which planetary nebulae arise are denoted as either \lq\lq young" or \lq\lq old".  Stellar populations with ongoing star formation are deemed young (disks of spirals, dwarf irregulars) while stellar populations where star formation has ceased are considered old (ellipitcals, bulges of spirals, dwarf spheroidals).  Though crude, this definition is sufficient for the purposes required.  

\begin{table}\def~{\hphantom{0}}
  \begin{center}
  \caption{Data Sources}
  \label{table_datasrc}
  \begin{tabular}{ll}\hline
      Galaxy  & PN Spectroscopy reference \\\hline%\\%[3pt]
       Fornax    & \cite{kniazevetal2006}\\
       Leo A & \cite{vanzeeetal2006}\\
       LMC & many sources, see \cite{stasinskaetal1998}\\
       M31   & \cite{jacobyford1986}, \cite{jacobyciardullo1999}, \cite{richeretal1999},\\
         & \cite{rothetal2004}\\
       M32   & \cite{richeretal1999}, \cite{richermccall2006b}\\
       NGC 147   & \cite{goncalvesetal2006}\\
       NGC 185   & \cite{richermccall2006b}\\
       NGC 205   & \cite{richermccall2006b}\\
       NGC 3109 & \cite{penaetal2006}\\
       NGC 4697 & \cite{mendezetal2005}\\
       NGC 5128 & \cite{walshetal1999}\\
       NGC 6822 & \cite{richermccall2006a}\\
       Sagittarius & \cite{walshetal1997}, \cite{zijlstraetal2006}\\
       Sextans A & \cite{magrinietal2005}, \cite{kniazevetal2005}\\
       Sextans B & \cite{magrinietal2005}, \cite{kniazevetal2005}\\
       SMC & many sources, see \cite{stasinskaetal1998}\\
       \hline
  \end{tabular}
 \end{center}
\end{table}

\section{Spectral Properties}\label{spec_props}

Given the constancy of at least the high luminosity part of the planetary nebula luminosity function, it is useful to inquire whether the bright planetary nebulae in all galaxies are similar.  Ideally, the properties of both the nebular shells and the central stars should be compared for planetary nebulae from different galaxies.  \cite{stasinskaetal1998} is adopted as a guide: the central star temperature will be characterized via the $\mathrm{He~II}\lambda4686/\mathrm H\beta$ ratio, while the nebular ionization structure will be characterized using the $\mathrm{[O~III]}\lambda5007/\mathrm{[O~II]}\lambda3727$ ratio.   

\begin{figure}
 \begin{center}
 \includegraphics[angle=-90,width=0.8\linewidth]{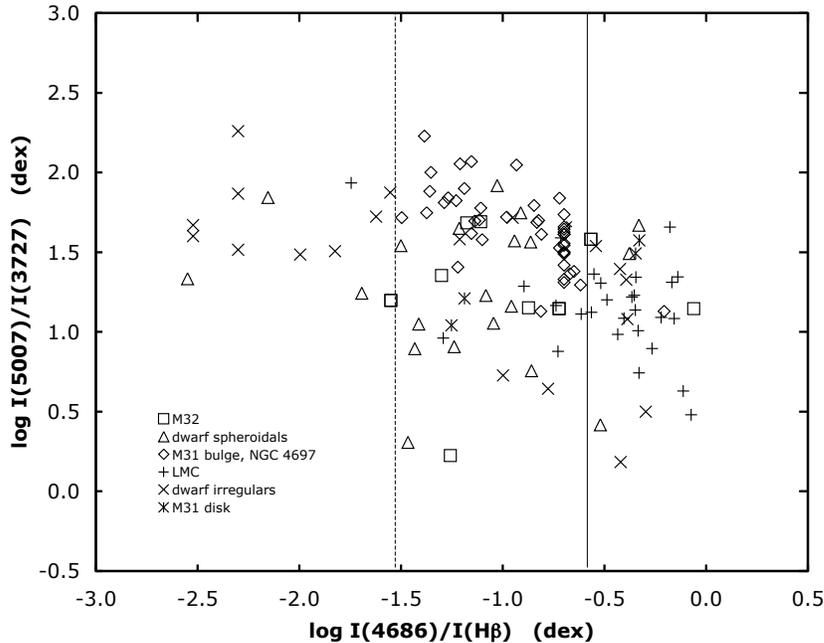}
  \caption{Here, the central star temperatures ($\mathrm{He~II}\lambda4686/\mathrm H\beta$) and nebular ionization structures ($\mathrm{[O~III]}\lambda5007/\mathrm{[O~II]}\lambda3727$) are compared for planetary nebulae from different galaxies.  The planetary nebulae in NGC 4697 produce the vertical string of diamonds, since their $\mathrm{He~II}\lambda4686/\mathrm H\beta$ ratios are all upper limits.  Here and in subsequent figures, the dwarf spheroidal symbols represent objects in Fornax, NGC 185, NGC 205, and Sagittarius, while the dwarf irregular symbols represent objects in Leo A, NGC 3109, NGC 6822, Sextans A, Sextans B, and the SMC. In Centaurus A (NGC 5128), only a small fraction of planetary nebulae are found to the right of the solid line \cite{rejkubawalsh2004}.
  }\label{figure_props}
 \end{center}
\end{figure}

Fig. \ref{figure_props} presents the $\mathrm{[O~III]}\lambda5007/\mathrm{[O~II]}\lambda3727$ ratio as a function of the $\mathrm{He~II}\lambda4686/\mathrm H\beta$ intensity ratio.  At high $\mathrm{He~II}\lambda4686/\mathrm H\beta$ ratios, almost all of the planetary nebulae arise from young stellar populations.  At intermediate $\mathrm{He~II}\lambda4686/\mathrm H\beta$ ratios, the planetary nebulae from old stellar populations dominate.  For $\mathrm{He~II}\lambda4686/\mathrm H\beta < 0.03$, $\mathrm{He~II}\lambda4686$ is very difficult to detect with existing spectroscopy.  These distributions reflect different temperatures distributions for the central stars in planetary nebulae from young and old stellar populations.  There are also systematic changes in $\mathrm{[O~III]}\lambda5007/\mathrm{[O~II]}\lambda3727$ among galaxies, with larger values occurring in old progenitor populations.  As \cite{stasinskaetal1998} advocate, the sequence observed in Fig. \ref{figure_props} is accidental.  Models generally predict an evolution from the upper right to lower left, contrary to the trend defined globally in Fig. \ref{figure_props}, though the exact tracks depend upon model details.  To reproduce the distribution observed in Fig. \ref{figure_props} it is necessary to adopt different models and different distributions of model parameters for the planetary nebulae in different galaxies.  Both the central stars (temperature distributions) and nebular shells (the distribution of ionization structures) must differ among the bright planetary nebulae in different galaxies.  

The different distributions of properties for planetary nebulae in different galaxies imply that bright extragalactic planetary nebulae do not constitute a single, homogeneous population.  It is likely these differences that explain the residual dependence upon progenitor age that permit planetary nebulae from old stellar populations to produce up to twice the [O~III]$\lambda$5007 luminosity of their counterparts in young stellar populations.  Otherwise, the maximum [O~III]$\lambda$5007 luminosity depends upon oxygen abundance (\cite[Dopita \etal\ 1992]{dopitaetal1992}; \cite[Richer \& McCall 1995]).  

\section{Chemical Abundances}

While nucleosynthetic processing in their stellar progenitors modifies some of the elemental abundances observed in planetary nebulae, others are unaffected.  The former inform us of stellar evolution while the latter may be used to probe galactic evolution.

\begin{figure}
 \begin{center}
 \includegraphics[angle=-90,width=0.8\linewidth]{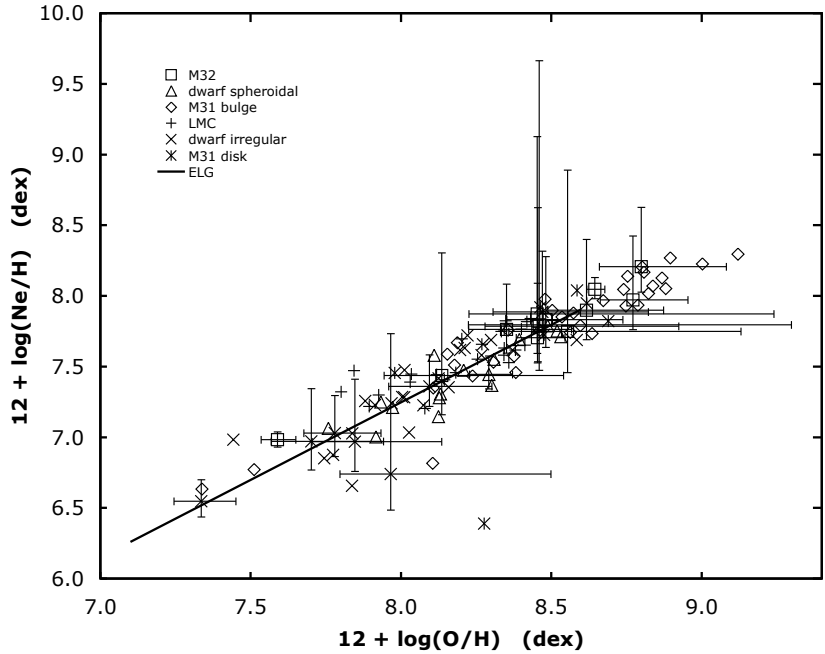}
  \caption{Here, the neon abundances are plotted as a function of the oxygen abundances.  For clarity, error bars are only shown for Leo A, M32, and NGC 3109, which are representative of the range of uncertainty for all of the data.  The solid line is the relation between neon and oxygen abundances in emission line galaxies from \cite{izotovetal2006}.
  }\label{figure_one}
 \end{center}
\end{figure}

Fig. \ref{figure_one} presents the relation between neon and oxygen abundances in bright extragalactic planetary nebulae.  Clearly, there is an excellent correlation between the abundances of these two elements for the large majority of objects.  Furthermore, the correlation observed is in excellent agreement with the relation found in emission line galaxies (ELGs, \cite[Izotov \etal\ 2006]{izotovetal2006}), where the neon and oxygen abundances are set by the nucleosynthetic yields of type II supernovae.  That the same relation is found for bright planetary nebulae is most simply explained if their stellar progenitors normally do not modify either abundance.  Nonetheless, a few rare planetary nebulae from young, low metallicity stellar populations have dredged up oxygen (e.g., the planetary nebula in Sextans A).  It appears, however, that oxygen dredge up is not a common phenomenon.  

\begin{figure}
 \begin{center}
 \includegraphics[angle=-90,width=0.8\linewidth]{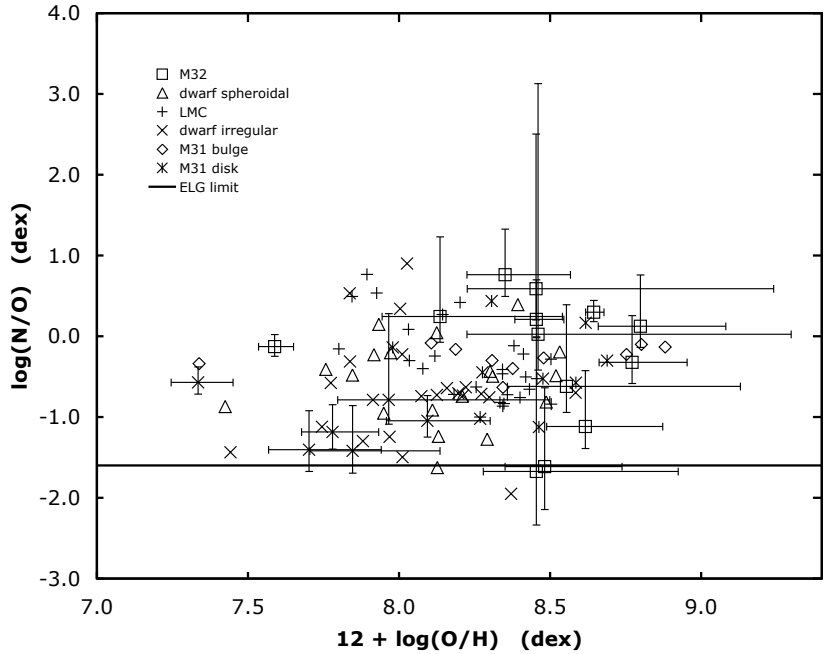}
  \caption{Here, the nitrogen abundances are plotted as a function of the oxygen abundances.  Again, error bars are only shown for Leo A, M32, and NGC 3109, which are representative of the range of uncertainty for all of the data.  The solid line is the lower limit to the N/O ratio observed in emission line galaxies from \cite{izotovetal2006}.  
  }\label{figure_on}
 \end{center}
\end{figure}

Fig. \ref{figure_on} presents N/O abundance ratio as a function of oxygen abudance.  Nitrogen enrichment is expected to be a function of the progenitor mass, e.g., \cite{marigo2001}.  If so, for a given oxygen abundance, one expects larger nitrogen enrichment in planetary nebulae arising from young stellar populations, a result that is not observed.  Instead, a similar range of nitrogen enrichment is found in planetary nebulae from all stellar populations.  While surprising, similar evidence of nitrogen enrichment in old stellar populations has been found previously for M32 (\cite[Stasi\'nska \etal\ 1998]{stasinskaetal1998}) and all recent studies of the Milky Way bulge (\cite[Cuisinier \etal\ 2000]{cuisinieretal2000}; \cite[Escudero \& Costa 2001]{escuderocosta2001}; \cite[Escudero \etal\ 2004]{escuderoetal2004}; \cite[Exter \etal\ 2004]{exteretal2004}; \cite[G\'orny \etal\ 2004]{gornyetal2004}).  In M32, planetary nebulae span the full range of nitrogen enrichment at a given oxygen abundance, while, in the bulge of M31, planetary nebulae with similar nitrogen enrichment are found spanning more than a decade in oxygen abundance, both of which argue that this nitrogen enrichment results from the evolution of the stellar progenitors rather than the chemical evolution of their host galaxies.  While nitrogen enrichment appears to be random in old stellar populations, it is not clearly a function of progenitor mass in young stellar populations, since a positive correlation between N/O and oxygen abundance is not necessarily seen for individual galaxies.  Fig. \ref{figure_on} does not preclude a tendency for larger nitrogen enrichment at higher progenitor masses, however, if these more massive progenitors do not produce bright planetary nebulae.

Finally, the helium abundances in bright extragalactic planetary nebulae occasionally appear to depend upon the chemical evolution of their host galaxy (\cite[Richer \& McCall 2006c]{richermccall2006c}).  At a given oxygen abundance, the planetary nebulae in dwarf spheroidals, and perhaps in M32, tend to have systematically higher helium abundances than do planetary in other galaxies.  Since helium enrichment should accompany nitrogen enrichment and since nitrogen is not systematically enriched in dwarf spheroidals, it seems likely that the tendency for high helium abundances in dwarf spheroidals is attributable to the chemical evolution of these galaxies.

\section{Kinematics}

\begin{figure}
 \begin{center}
 \includegraphics[angle=-90,width=0.8\linewidth]{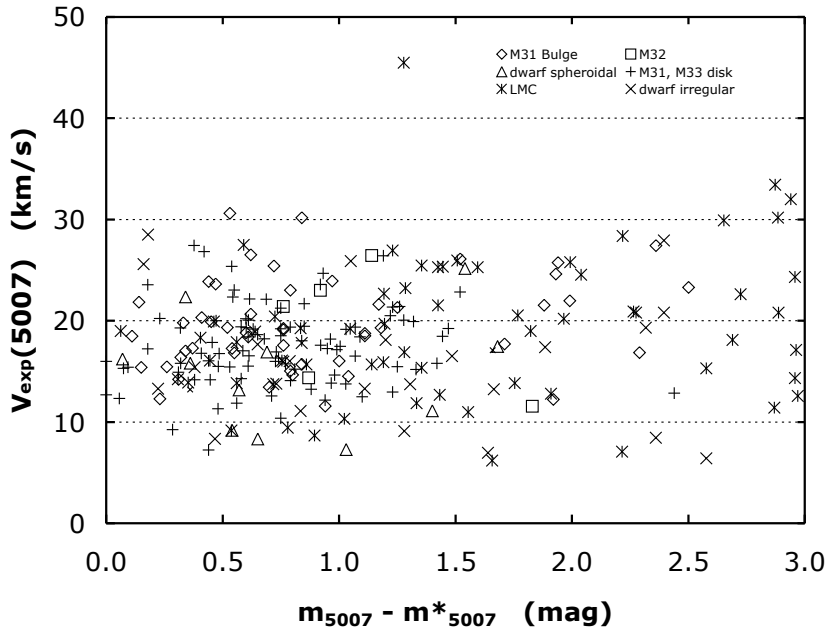}
  \caption{The expansion velocity in [O~III]$\lambda$5007 is plotted as a function of the [O~III]$\lambda$5007 luminosity relative to the peak of the luminosity function.  Clearly, the expansion velocity does not depend upon [O~III]$\lambda$5007 luminosity (as expected: e.g., \cite[Stasi\'nska \etal\ 1998]{stasinskaetal1998}).  There is no obvious segregation of the planetary nebulae from young and old stellar populations.
  }\label{figure_vexplo3}
 \end{center}
\end{figure}

Recently, line profiles in [O~III]$\lambda$5007 have been  obtained for over 170 planetary nebulae in a variety of Local Group galaxies at the Observatorio Astron\'omico Nacional in San Pedro M\'artir, Mexico (e.g., \cite[L\'opez \etal\ 2006]{lopezetal2006}).  Apart from these data, the only extant kinematic data are for planetary nebulae in the Magellanic Clouds (\cite[Dopita \etal\ 1986]{dopitaetal1985}; \cite[Dopita \etal\ 1988]{dopitaetal1988}) and the Sagittarius dwarf spheroidal (\cite[Zijlstra \etal\ 2006]{zijlstraetal2006}).  In all cases, data are available in [O~III]$\lambda$5007, and occasionally in other lines.  Given the distances, these objects are usually point sources, the exceptions being some of the planetary nebulae in Sagittarius and the Magellanic Clouds, so the line profiles are generally spatially unresolved.  Except for the brightest few objects, the line profiles are well-described by a gaussian shape.  Similar observations of planetary nebulae in the Milky Way bulge indicate that the gaussian line profiles are due to the lower signal-to-noise achieved for most of the extragalactic planetary nebulae, but, even in the Milky Way bulge, the departures from a gaussian line shape typically represent less than 10\% of the total flux.  Where it has been possible to compare, the line widths in [O~III]$\lambda$5007 and H$\alpha$ are very similar, so the kinematics from [O~III]$\lambda$5007 appear to be representative of most of the mass in bright planetary nebulae (\cite[B\'aez 2006]{baez2006}; \cite[L\'opez \etal\ 2006]{lopezetal2006}).  Half of the FWHM is adopted as a measure of the mean expansion velocity, though, given the gaussian line shapes, this may be easily converted to velocities at any other intensity level.  

Fig. \ref{figure_vexplo3} illustrates that the expansion velocities of bright planetary nebulae show no strong dependence upon either the age or metallicity of the progenitor stellar population (see also \cite[L\'opez \etal\ 2006]{lopezetal2006}).  In Fig. \ref{figure_vexplo3}, planetary nebulae from stellar populations with and without ongoing star formation span metallicities from the solar value to under one tenth of this value, e.g., from the disk and bulge of M31 to dwarf irregulars and spheroidals.  

Fig. \ref{figure_vexplo3} emphasizes the youth of bright extragalactic planetary nebulae, since their expansion velocities overlap the range of values found for AGB envelopes.  Later, energy input from the central stars is expected to dominate the kinematics(\cite[Villaver \etal\ 2002]{villaveretal2002}).  The line profile shapes also reflect their youth: At early times, hydrodynamical models predict considerable material at low velocities within the bright main shell (\cite[Villaver \etal\ 2002]{villaveretal2002}; \cite[Perinotto \etal\ 2004]{perinottoetal2004}) that should \lq\lq fill in" the flat-topped profile of an unresolved shell (e.g., \cite[Gesicki \& Zijlstra 2000]{gesickizijlstra2000}).  

\begin{figure}
 \begin{center}
 \includegraphics[angle=-90,width=0.8\linewidth]{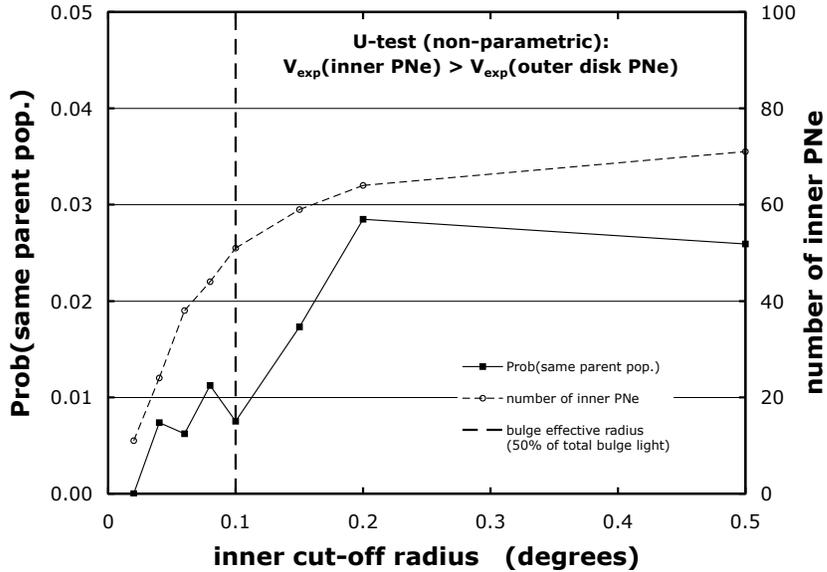}
  \caption{Here, the expansion velocity distribution for the 29 planetary nebulae in M31's outer disk are compared with that for different samples of planetary nebulae in M31's bulge and inner disk.  The left vertical axis is the probability that the two samples are drawn from the same parent distribution (based upon a U-test) while the right vertical axis shows the number of planetary nebulae in the inner sample.  
  }\label{figure_utest}
 \end{center}
\end{figure}

The foregoing notwithstanding, the kinematics of planetary nebulae in M31 differ for old and young progenitor populations.  In Fig. \ref{figure_utest}, a U-test is used to compare a sample of 29 planetary in M31's outer disk (\cite[Richer \etal\ 2004]{richeretal2004}) with various samples defined at smaller radii.  (The U-test is basically a non-parametric version of the t-test for comparing the means of two distributions (\cite[Wall \& Jenkins 2003]{walljenkins2003}).)  Clearly, the distribution of expansion velocities for the outer disk planetary nebulae is significantly different from that for those in the bulge (within a radius of 0.1$^{\circ}$).  What is surprising is that the bulge planetary nebulae have slightly larger expansion velocities, by $\sim 3\,\mathrm{km}\,\mathrm s^{-1}$, a result that is significant at 99\% confidence.  Given the similar metallicities of M31's bulge and disk planetary nebulae (Table \ref{table_datasrc}), the difference in kinematics appears to be due to age alone.  If all of the planetary nebulae from old and young stellar populations are compared, at 97\% confidence, the planetary nebulae from old stellar populations have larger expansion velocities than their counterparts from young progenitor populations.   This may not be so unusual, since the models of \cite{villaveretal2002} indicate that expansion velocities for bright, young planetary nebulae are not a monotonic function of progenitor mass, thereby allowing larger expansion velocities from older progenitor populations.  

There is no correlation between expansion velocity and the nebular H$\beta$ luminosity, except for a lack of large expansion velocities at the highest luminosities (the youngest objects).  Expansion velocity correlates weakly with the [O~III]$\lambda$5007/H$\beta$ ratio.  Both quantities increase in step until the [O~III]$\lambda$5007/H$\beta$ ratio saturates, at which point the expansion velocity continues to increase somewhat.  The relation with [O~III]$\lambda$5007/H$\beta$ may have a dependence upon the age of the progenitor population, suggesting perhaps a generalization of the \cite{dopitaetal1988} results, wherein nebular kinematics are a function of the central star properties.

\begin{acknowledgments}

I thank all those who contributed data or information, often prior to publication, to allow me to make a more representative review of the properties of bright extragalactic planetary.  Those whose efforts are represented in this respect are M. Argote, A. S.-H. B\'aez, R. Calder\'on, R. L. M. Corradi, E. D\'\i az, D. R. Gon\c{c}alves, E. K. Grebel, N. Hwang, A. Y. Kniazev, M. G. Lee, P. Leisy, G. L\'opez, J. A. L\'opez, M. L. McCall, L. Magrini, A. Mampaso, M. Pe\~na, A. G. Pramskij, S. A. Pustilnik, M. Rejkuba, H. Riesgo, D. Rocha, G. Stasi\'nska, and J. R. Walsh.  I gratefully acknowledge financial support from UNAM DGAPA grants IN108406-2, IN108506-2, and IN112103 as well as CONACyT grant 43121.

\end{acknowledgments}

\end{document}